# Enhancing Medical Imaging with GANs Synthesizing Realistic Images from Limited Data

Yinqiu Feng[1], Bo Zhang[2], Lingxi Xiao[3], Yutian Yang[4], Tana Gegen[5], Zexi Chen[6]

[1]Columbia University, USA

[2]Texas Tech University, USA

[3]Georgia Institue of Technology, USA

[4]University of California, Davis, USA

[5]Johns Hopkins University, USA

[6]North Carolina State University, USA

*Abstract: In this research, we introduce an innovative method for synthesizing medical images using generative adversarial networks (GANs). Our proposed GANs method demonstrates the capability to produce realistic synthetic images even when trained on a limited quantity of real medical image data, showcasing commendable generalization prowess. To achieve this, we devised a generator and discriminator network architecture founded on deep convolutional neural networks (CNNs), leveraging the adversarial training paradigm for model optimization. Through extensive experimentation across diverse medical image datasets, our method exhibits robust performance, consistently generating synthetic images that closely emulate the structural and textural attributes of authentic medical images.*

## I. INTRODUCE

With the continuous development of imaging technologies such as MRI and ultrasound, doctors can more accurately understand their patients' conditions and provide more effective treatment options. However, access to medical imaging data is often limited because it requires expensive equipment and specialized technology to produce, and is also subject to privacy and ethical limitations. Consequently, utilizing the limited medical image data to generate more diversified and rich image data has become an important challenge in the field of medical imaging. Traditional methods typically rely on hand-designed feature extractors or statistical model-based approaches. However, these methods often struggle to capture complex nonlinear relationships and advanced features in image data, resulting in synthetic images that lack authenticity and diversity.

GAN(Generative Adversarial networks), as a powerful deep learning framework, has made remarkable achievements in the field of image generation[1]. GANs consists of generator and discriminator. By means of adversarial training, the sample generated by the generator is as realistic as possible to deceive the discriminator. This competitive process forces the generator to continuously improve the quality of the generated samples, eventually reaching a level similar to the real sample.

In the field of medical imaging, use GANs to generate realistic medical images has become a concerned research direction[2-4]. The generated synthetic medical images can be used not only for data enhancement to improve the generalization ability of machine learning models, but also for medical image synthesis[5], data enhancement, pathological analysis[6-8], and medical prediction[9]. However, due to the particularity and complexity of medical image data[10], the direct application of traditional GANs model to medical image synthesis often faces some challenges, such as data scarcity, sample imbalance, and integration of medical knowledge.

This study aims to address the scarcity and lack of diversity of medical image data, and proposes a new medical image synthesis method based on generative adversarial networks (GANs). Our approach constructs generators and discriminators through deep convolutional neural networks, and employs adversarial training strategies to continuously optimize model parameters to generate realistic synthetic medical images. Compared with traditional medical image synthesis methods, our method has several significant advantages:

First, our method is able to generate a large number of synthetic images from a small amount of real medical image data. Through our method, we can effectively expand the medical image data set and improve the richness and diversity of data.

Secondly, the synthesized images we generate are able to maintain structural and textural features similar to the real images. This means that our model is not only able to generate realistic synthetic medical images, but the resulting images have visual features and anatomical structures similar to real images. This is important for medical image analysis and diagnosis, because doctors rely on image data to make accurate diagnosis and treatment decisions, and the authenticity and reliability of synthetic images are key to their acceptance in clinical practice.

The experimental findings indicate that our proposed approach exhibits strong performance across various medical image datasets. The synthesized images not only possess a high degree of visual realism but also demonstrate practical utility and interpretability in medical contexts. This presents a novel and effective avenue for medical image synthesis, poised to advance the realms of medical image analysis and diagnosis by furnishing clinicians with valuable supplementary insights. Moreover, it is anticipated that this methodology will catalyze further advancements and applications within the domain of medical image technology.

Moving forward, our research endeavors will focus on refining our techniques to enhance both the quality and diversity of generated medical images. We aim to delve into more intricate model architectures and training methodologies to effectively tackle the intricate and diverse nature of medical imaging data. Furthermore, we will explore the practical applications of the generated synthetic images in clinical settings to validate their feasibility and efficacy in medical image diagnosis and treatment. Through ongoing refinement and optimization efforts, we envision that the medical image synthesis approach leveraging generative adversarial networks will engender significant breakthroughs and advancements in the realm of medical imaging.

## II. RELATED WORK

Image super-resolution technology, which enhances low-resolution images to high resolution, is a pivotal domain within computer vision with extensive applications, including medical imaging and surveillance[11-12]. This paper concentrates on its utilization in medical imaging, with the dual objectives of augmenting visual quality and refining computer vision task performance. The inherent challenge lies in the fact that low-resolution images correspond to multiple potential high-resolution solutions. Various techniques are employed to address this, encompassing interpolation[13], prediction, edge classification[14-15], statistics, image block processing, and sparse representation. However, traditional methods entail manual feature extraction and intricate reconstruction processes, posing operational complexities [16].

Algorithms based on feature matching strive to map similar regions between the original and target modalities. They can typically be categorized into four types: segmentation-based, sparse coding-based, patch-based, and atlas-based[17-18]. The segmentation-based approaches initially segment MRI images into components like soft tissue, air, or bone. Subsequently, these components are aligned with analogous structures in CT images, employing methods like fuzzy clustering to achieve precise segmentation[19]. However, such methods may be at risk of failure as different tissue regions share predefined CT values and ambiguous class classifications, such as in cases of air and bone. The method based on sparse coding first extracts patches from the original MRI image, then encodes them using a patch dictionary constructed using a linearly registered MRI atlas, and converts the obtained sparse encoding into a CT constructed using the registered CT atlas. Patch dictionary. Finally, the synthesized CT patches are converted into target CT images. However, sparse coding needs to be optimized on all image areas, resulting in high computational costs, and building a global dictionary also increases the time to solve sparse coding. Patch-based methods attempt to estimate the nonlinear mapping between two modality images, obtain patches of existing modality images, and then estimate the center pixel intensity corresponding to the patch of the target modality image[20]. Atlas-based methods mainly utilize a collection of image pairs collected from source modality images and target modality images to learn predictions between cross-modal images.

SRCNN [21] innovation not only allows training using existing image datasets, but also improves the ideal degree of reconstruction. The overall framework includes three convolutional layers of different sizes for feature extraction and one deconvolution layer for image amplification, providing an important reference for subsequent work on super resolution.

In 2017, Dong Nie, Roger Trullo, and their colleagues introduced a groundbreaking 3D generative adversarial network (GAN) model employing a full convolutional network (FCN) as a generator. This model aimed to convert brain magnetic resonance imaging (MRI) data into computed tomography (CT) images [22]. The 3D approach surpassed traditional 2D processing by comprehensively capturing spatial information of the brain and minimizing discontinuities between image slices. During model training, a discriminator and image gradient differences were integrated to ensure the quality and sharpness of the generated CT images[23]. Additionally, the Auto-

Context model iteratively optimized context information[24]. Application of this method to ADNI public and private pelvic datasets demonstrated its efficacy in predictive CT imaging.

### III. GENERATIVE ADVERSARIAL NETWORK

In 2014, Goodfellow and his collaborators introduced the concept of generative adversarial networks (GANs)[25], which has been hailed as a groundbreaking advancement. Since its inception, GANs have spurred a surge of research activity, with numerous papers published annually, underscoring their profound impact and versatility across diverse domains, including computer vision, speech recognition, and natural language processing. Rooted in the principles of zero-sum game theory, the GAN model consists of two components: a generator and a discriminator[26]. The generator is tasked with producing synthetic data samples by learning from random noise inputs, aiming to replicate a specific data distribution. Conversely, the discriminator's role is to differentiate between real and synthetic data, assigning probabilities to input samples to ascertain their authenticity. The conceptual framework of a generative adversarial network is elucidated in Figure 1.

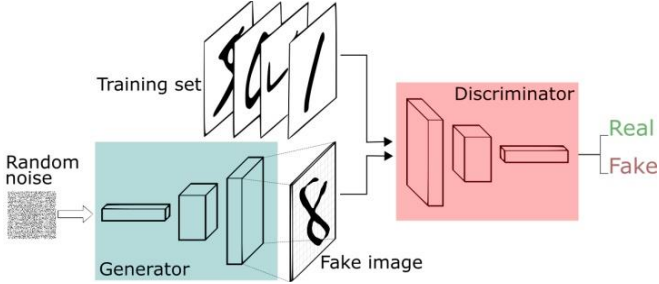

Figure 1 Generate the adversarial network model schematic

Figure 2 Generate the adversarial network model schematic

$$G = \arg\min_G Div(P_z(z), P_{data}(x)) \qquad (1)$$

Where Div(·) represents the difference between the two distributions, z represents the input noise data, which follows the distribution $P_z$, and $x$ real data follows the $P_{data}$ distribution. Discriminator D is used to classify the input data, that is, to judge whether the input data belongs to the real data or the generated data. Then the optimization objective of discriminator D is as follows:

$$T = \arg\max_T V(G, R) \qquad (2)$$

The final optimization goal of GAN network can be defined as:

$$\min_G \max_T E_{x \sim P_{data}}[\log T(x)] + E_{z \sim P_z}[\log(1 - T(G(z)))] \qquad (3)$$

In the optimization objective function of GAN, the discriminator needs to make DG(z) as close to 0 as possible, while the generator needs to generate high-quality samples with the same distribution as the real sample so that DG(z) is as close to 1 as possible. The two networks are alternately trained by gradient descent, and when the performance of the generator and discriminator is trained well enough, the discriminator's output is close to 0.5 for all inputs, thus achieving an equilibrium state.

Nevertheless, traditional GAN training encounters issues of instability including gradient vanishing and model collapse. To address these issues, Gulrajani et al. introduced the Wasserstein Generative Adversarial Network (WGAN). Contrary to the original GAN that utilizes JS-KL divergence to derive the optimal solution for the generator and discriminator, WGAN employs the Wasserstein distance as its objective function and enhances this function by optimizing the discriminator. The optimization objectives for the generator and discriminator are outlined below:

$$G = \arg\min_G E_{x \sim P_r}\left[T_w(x)\right] - E_{\tilde{x} \sim P_g}\left[T_w(\tilde{x})\right] \qquad (4)$$

$$T = \arg\max_D E_{x \sim P_r}\left[T_w(x)\right] - E_{\tilde{x} \sim P_g}\left[T_w(\tilde{x})\right] \qquad (5)$$

Where $P_r$ represents the real data distribution, $P_g$ represents the generated data distribution, and w represents the truncation parameter.

Here, $P_r$ denotes the actual data distribution, $P_g$ signifies the distribution of the generated data, and $W$ represents the truncation parameter.

### IV. KNOWLEDGE DISTILLATION

Knowledge distillation (KD) achieves model compression by compressing the performance of the teacher network into the student network, as shown in Figure 2. KD trains the teacher network, takes its soft labels and hard labels, and migrates those labels to the student network to accomplish the same task. KD is divided into three steps: training teacher model, building student model and knowledge transfer. In the knowledge transfer stage, soft label loss and hard label loss are used to optimize the student network. Soft labels are obtained from the softmax output of the teacher model, and hard labels are the labels of the real data. After the input of real data, the soft prediction label is obtained through the softmax layer of the student network, and the hard prediction label is obtained through the softmax function with T-value of 1. The T-value is used to narrow the distribution differences, and when T=1, the softmax output is the same as when the T-value is not considered. Increasing T-value smooths the probability distribution of softmax output and increases the amount of information. The output of the softmax layer can be defined as:

$$q_i = \frac{\exp(z_i / T)}{\sum_j \exp(z_j / T)} \qquad (6)$$

Where $q_i$ represents the probability of softmax layer output and $z_i$ represents the original probability distribution of input. In the process of knowledge distillation, distillation losses need to be used to update and train the student network. The distillation loss is the weighted sum of the soft and hard losses. Distillation loss L is defined as:

$$L = \alpha L_{soft} + \beta L_{hard} \qquad (7)$$

$L_{soft}$ and $L_{hard}$ indicate the cross-entropy loss of soft labels and hard labels.

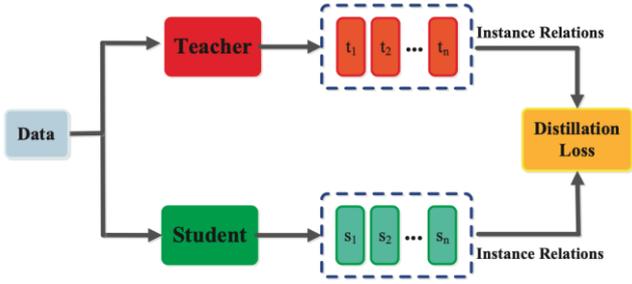

Figure 3 Knowledge distillation process

## V. GENERATIVE ADVERSARIAL NETWORK OF FUSION KNOWLEDGE DISTILLATION

The framework of KGAN, illustrated in Figure 3, begins by training a high-performing teacher GAN model on a large dataset of medical images to produce high-fidelity medical images. Following this, we construct a lightweight student GAN model, which bears resemblance to the teacher model but with a reduced parameter count. Knowledge transfer from the teacher to the student model occurs through the utilization of both soft labels, which are derived from the probability distribution of images generated by the teacher model, and hard labels, obtained from real medical images. Throughout the training process, the student model endeavors to minimize both soft label loss and hard label loss, gradually converging towards the performance level of the teacher model. This innovative approach efficiently transfers the intricate knowledge encoded within the teacher model to the student model in medical image synthesis tasks, resulting in reduced computational and storage overheads without compromising on the quality of the synthesized images.

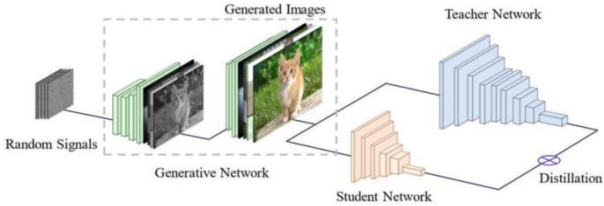

Figure 4 KGAN frame

## VI. EXPERIMENT

### A. Data set

The experimental dataset utilized in this study was sourced from Harvard Medical School, USA, accessible through their website (http://www.med.harvard.edu). We meticulously selected 300 pairs of high-quality CT and MR images exhibiting clear brain texture and abundant detail features from their openly available common brain disease image dataset. The training set comprised 200 pairs of images dedicated to network training, while the test set consisted of 100 pairs of images used to assess the network's generalization performance. The dataset utilizes Linked Data techniques to merge diverse data formats, improving data interoperability and analysis across domains such as machine learning and AI[27]. This strategy aids in dismantling data silos, augments dataset variety, and fosters the creation of more precise and scalable AI solutions. To mitigate potential issues of network overfitting arising from the dataset's relatively small size, we standardized the MR and corresponding CT images to a size of 256×256 pixels[28]. Additionally, we implemented random flipping to augment the dataset's diversity, thereby enhancing the robustness and generalization capability of the trained network.

### B. Model optimization

The loss Settings for the generator are as follows:

$$L_G = -\frac{1}{m}\sum_{i=1}^{m}\log D(G(z^{(i)})) \qquad (8)$$

The discriminator loss is set as follows:

$$L_D = -\frac{1}{m}\sum_{i=1}^{m}\log D(x^{(i)}) - \frac{1}{m}\sum_{i=1}^{m}\log(1 - D(G(z^{(i)}))) \quad (9)$$

### C. Evaluation index

To objectively assess the fusion effect, three evaluative metrics were employed to examine the efficacy of KGAN on the brain medical image dataset, encompassing spatial frequency (SF), structural similarity index (SSIM), and difference-dependent sum (SCD). Spatial frequency (SF) quantifies the resolution of the fused image, with higher SF values suggesting greater clarity of details. The structural similarity index (SSIM) appraises the image's brightness, contrast, and structural congruence, where higher SSIM scores indicate enhanced structural integrity. The difference-dependent sum (SCD) evaluates the extent to which the source image is preserved in the fused image by assessing the discrepancies between the fused and the original source images. A substantial and positive SCD value denotes a more pronounced correlation between the fused image and source image A.

### D. Result analysis

Table I Experimental results under different methods

| Method | SF | SSIM | SCD |
|---|---|---|---|
| CycleGAN | 15.31 | 0.542 | 1.256 |
| CSGAN | 15.52 | 0.558 | 1.197 |
| KGAN | 16.37 | 0.593 | 1.339 |

Table I compares the performance of three methods - CycleGAN, CSGAN, and KGAN - on the brain medical image fusion task. Spatial frequency (SF), structural similarity index (SSIM) and differential correlation sum (SCD) were evaluated. KGAN has the best performance on all indexes. Its spatial frequency (16.37) indicates that its fusion image is the clearest in detail, and its structural similarity index (0.593) indicates that its fusion image is the closest to the original image in brightness, contrast and structure retention. The sum of difference correlation (1.339) indicates that it has the best effect on preserving source image information. In contrast, CSGAN and CycleGAN performed slightly less well, especially when it came to retaining source image information, with CSGAN having the lowest SCD value (1.197), indicating a weaker ability to retain information. These data show that KGAN is more efficient at processing the fusion task of medical images of the brain.

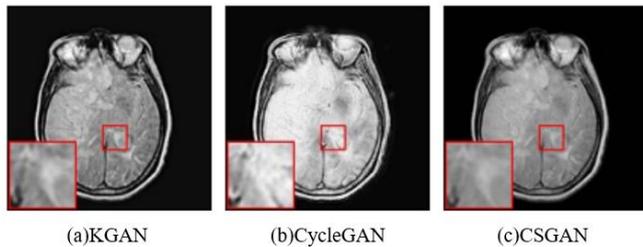

(a)KGAN　　　(b)CycleGAN　　　(c)CSGAN

Figure 4 Fusion experiment results of different methods

Figure 4 is the fusion results obtained from different models, namely CycleGAN, CSGAN, and KGAN. Notably, each model exhibits distinct characteristics in terms of image quality and fidelity. CycleGAN, while demonstrating high contrast, suffers from overexposure issues. This is evident in the loss of information retention from the original image, where certain details are obscured due to excessive brightness and contrast enhancement. On the other hand, CSGAN exhibits deficiencies in detail retention and suffers from oversmoothing problems, leading to a reduction in image clarity. The images generated by CSGAN lack fine-grained textures and appear overly smoothed, compromising the visual fidelity of the synthesized images. In contrast, KGAN achieves a more balanced outcome by effectively restoring local details while preserving the essential information present in the original image. The fusion results produced by KGAN exhibit improved clarity and detail retention compared to the other models, indicating its superior performance in synthesizing high-quality images. These findings underscore the importance of model selection and parameter tuning in achieving desirable image synthesis outcomes and highlight KGAN's efficacy in addressing the challenges associated with overexposure and oversmoothing.

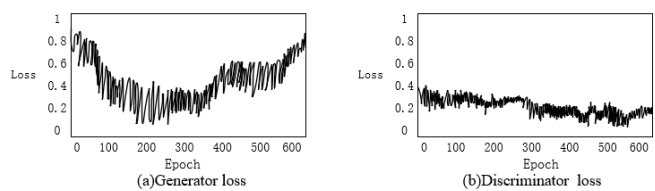

(a)Generator loss　　　(b)Discriminator loss

Figure 5 Generator and discriminator loss changes

Figure 5 illustrates the evolution of both the generator and discriminator loss values across epochs during the training process. Initially, there is considerable fluctuation in the loss values, which is a common occurrence in training deep learning models. However, as training progresses, the losses tend to stabilize, with the generator loss showing a slight upward trend after an initial decline. This phenomenon suggests a possible occurrence of overfitting, wherein the generator starts to memorize the training data rather than learning generalizable features. Conversely, the discriminator loss exhibits a smoother trajectory, gradually converging to a value around 0.2. This convergence indicates that the discriminator's performance becomes more stable over time, suggesting that it effectively learns to distinguish between real and synthetic images. Overall, these observations highlight the dynamics of the training process and underscore the importance of monitoring loss values to ensure the stability and effectiveness of the model.

## VII. CONCLUSION

In this article, a method for synthesizing medical images using generative adversarial networks (GANs) is introduced and validated. Utilizing a generator and discriminator built on deep convolutional neural networks, this approach is capable of producing high-quality synthetic images from a limited pool of real medical image data through the refinement of an adversarial training strategy. The structural and textural attributes of these synthetic images bear a strong resemblance to those of authentic images, demonstrating the method's exceptional generalization capabilities. The experimental findings indicate that this method is effective across various medical image datasets. The synthetic images not only broaden the diversity of the datasets but also enhance their utility, which holds substantial importance for medical research and clinical applications. These outcomes suggest the promising potential and expansive applicability of GANs in the realm of medical imaging.

Future research will focus on the following directions: First, we will focus on improving and optimizing GANS-based medical image synthesis methods, exploring more effective model structures and training strategies to improve the quality and diversity of synthesized images. Secondly, we will investigate how to further improve the generalization ability and adaptability of the model through techniques such as data enhancement and incremental learning to train more representative and robust models. In addition, we plan to further validate the feasibility and validity of the generated synthetic images in clinical practice and explore their impact on clinical decision making. In addition, we will explore issues such as multimodal medical image synthesis and the interpretability and safety of images to ensure the reliability and safety of the generated images for clinical applications.